\documentclass[
  a4paper,
  fontsize=11pt,
  ]{article}

\pdfoutput=1

\usepackage{pos_ILD}
\usepackage{graphicx}

\usepackage[utf8]{inputenx}
\usepackage[T1]{fontenc}

\usepackage[british]{babel}
\usepackage{csquotes}

\usepackage{subcaption}
\usepackage{import}
\usepackage{xspace}
\usepackage{graphicx}
\usepackage[shortcuts]{extdash}
\usepackage{siunitx}
\DeclareSIUnit \lightspeed {\text{{c}}}
\usepackage{xcolor}
\definecolor{linkblue}{HTML}{264772}
\usepackage{hyperref}
\hypersetup{
    colorlinks=true,
    linktocpage=true,
    linkcolor=linkblue,
    citecolor=linkblue,
    urlcolor=linkblue
}

\usepackage[noabbrev,capitalise]{cleveref}

\usepackage{wrapfig}

\graphicspath{ {../pictures/} }
\DeclareGraphicsExtensions{.pdf,.png,.jpg}

\newcommand{\dEdx}{\ensuremath{\mathrm{d}E/\mathrm{d}x}\xspace}

\title{Implementation, performance and physics impact of particle identification at Higgs factories}

\author*[a]{Ulrich Einhaus}
\author[b,c]{Matthew Basso}
\author[a]{Mikael Berggren}
\author[d,e]{Valentina Cairo}
\author[a,f]{Bohdan Dudar}
\author[a]{Jenny List}

\affiliation[a]{Deutsches Elektronen-Synchrotron DESY, Notkestr. 85, 22607 Hamburg, Germany}
\affiliation[b]{TRIUMF, Wesbrook Mall, Vancouver, Canada}
\affiliation[c]{Department of Physics, Simon Fraser University, University Drive W, Burnaby, Canada}
\affiliation[d]{SLAC National Accelerator Laboratory, 2575 Sand Hill Road, Menlo Park, California 94025-7015, USA}
\affiliation[e]{CERN, 1211 Geneva 23, Switzerland}
\affiliation[f]{Uni Hamburg Department of Physics, Luruper Chaussee 149, 22761 Hamburg, Germany}

\emailAdd{ulrich.einhaus@desy.de}

\abstract{
This work introduces the software tool Comprehensive Particle Identification (CPID).
It is a modular approach to combined PID for future Higgs factories and implemented in the Key4hep framework.
Its structure is explained, the current module library laid out and initial performance measures for the ILD detector as an example presented.
A basic run of CPID works already as well as the default full-simulation ILD PID reconstruction, but allows for an easy and convenient addition of more PID observables, improving PID performance in future analyses and high-level reconstruction, such as strange tagging.
}

\FullConference{The European Physical Society Conference on High Energy Physics (EPS-HEP2023)\\
 21-25 August 2023\\
Hamburg, Germany\\}

\begin{document}
\maketitle


\newpage
\section{Introduction}

While the particle physics community agrees that the next collider should be an e$^+$e$^-$ Higgs factory, it is not clear which accelerator technology should be used, where it could be built, or even which shape (linear or circular) it should have.
Given limited person power resources, it is however necessary to use common approaches wherever possible.
One such area is naturally software tools and the community has agreed on a common framework, Key4hep \cite{Key4hep}.
It is hence desirable to develop tools within this framework in order to allow studies on any future machine to utilise them and to use them in a coherent way.
In addition, in recent years, the interest in particle identification (PID) has significantly increased, not least with the advent of the circular collider proposals which aim to provide an unrivaled physics programme at the electroweak scale which needs excellent flavour tagging.
Unfortunately, the tools for PID are at different levels of performance and realism for the different detector reconstruction chains and not easily mutually applicable.
Comprehensive PID (CPID) aims to provide a tool that can be applied to any future detector in the Key4hep framework.
It is currently implemented as a Marlin processor in the MarlinReco package \cite{MarlinReco}, using the LCIO event data model.
CPID takes a modular approach to both the extraction of PID information from the reconstruction chain as well as to the training and inference models which are used to combine this PID information most effectively.
The following sections describe the structure of CPID, its current module library, a performance example using the ILD detector \cite{ILD_IDR} as well as an outlook for physics application.

\section{CPID Structure}

\begin{figure}[!hbt]
  \centering
    \includegraphics[width=.75\textwidth,keepaspectratio=true]{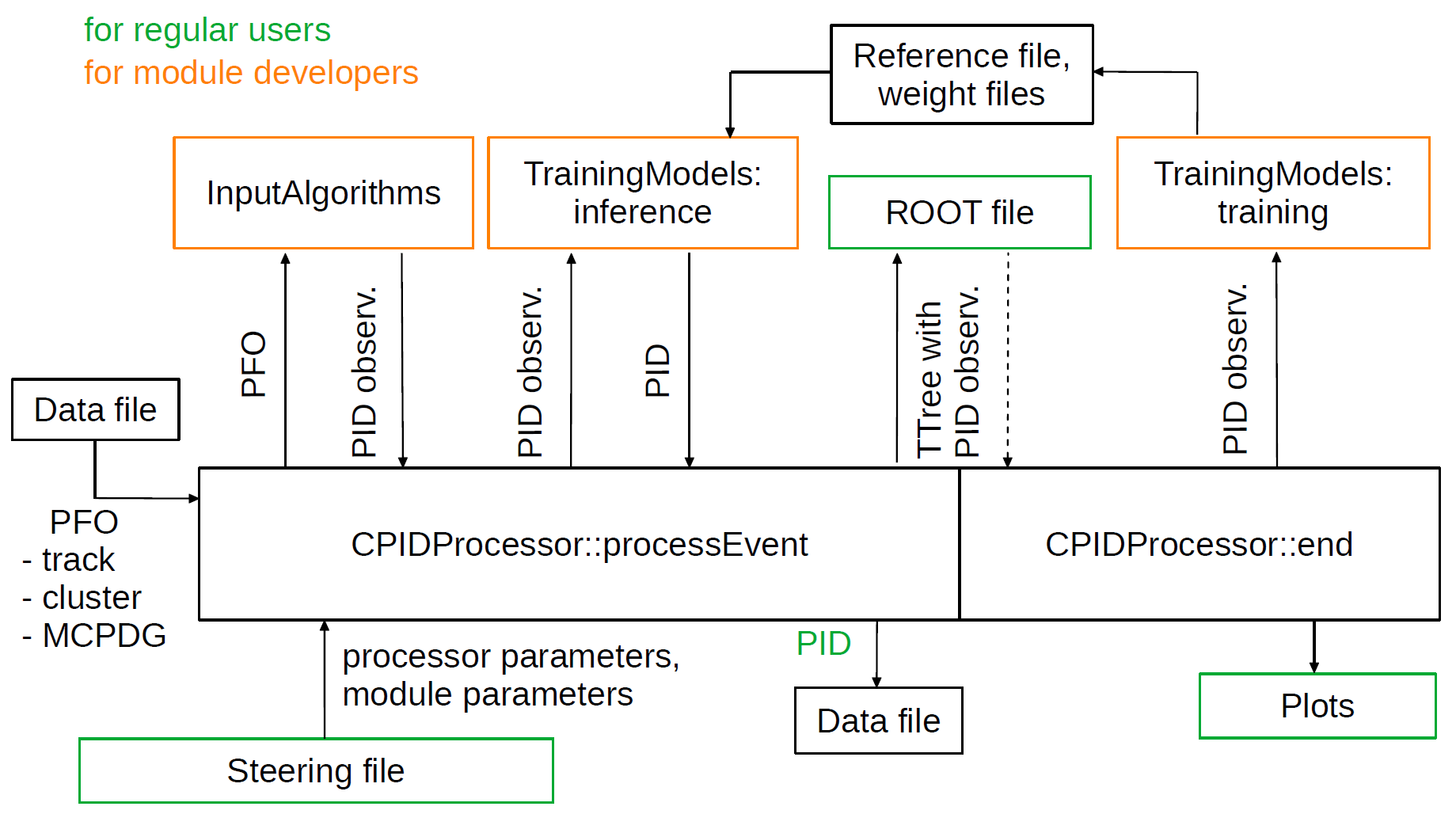}
    \caption{Schematic structure of the CPID functionality.}
  \label{fig:CPID}
\end{figure}

A schematic of the structure of CPID is shown in \autoref{fig:CPID}.
The core of CPID takes care of the central book keeping of PID observables and training.
From the input file, ParticleFlow objects (PFOs) are read, attached to which are relevant sub-objects such as tracks or calorimeter clusters, and - in simulation - also the Monte Carlo particle, which contains the true PID, referred to by its PDG number.
The PFOs are fed through the input algorithm modules, which each extract a specific PID observable or a set of these from the PFO.
In a training iteration, after all events have passed, their PID information is then given to the chosen training model module(s), which is/are trained.
A subdivision of the data can be made along the dimension of measured PFO momentum and a separate model instance is trained for each momentum bracket.
This approach can ease training, since PID observables are typically strongly dependent on momentum.
The output of the training are weight files along with a reference file that contains the set of chosen input algorithms, the momentum subdivision as well as the corresponding weight file names.
The observables can also be written out as a ROOT file, which allows to run separate training outside of Marlin.
In an inference iteration, a new set of PFOs is again fed through the input algorithms to extract their PID information, which is then directly used by the models with the weight files to determine the corresponding PDG likelihoods and potentially the best fitting PDG.
This information is added to the input file and can be used in the following, e.g.\ in flavour tagging or an analysis script.
With inference, CPID offers automatic performance plots to assess the chosen combination of modules.
The central book keeping as well as the modules are steered via a steering file, in which all parameters can be set.

As \autoref{fig:CPID} indicates in green, 'regular users', who want to use existing modules for extraction, training and inference, only need to adapt the steering file and can then use the resulting PID in the data file.
For external training, the ROOT file needs to be used, and it makes sense to have a look at the performance plots.
If a 'module developer'(orange) would like to add a new PID observables, i.e.\ a new input algorithm or a new training model, they need to provide a new such module and re-install the corresponding MarlinReco installation.
Both the input algorithm as well as the model modules have well-defined interfaces with the central book keeping core, which makes it convenient to fit in such new modules.
However, a module developer does not need to touch the CPID core and new modules are automatically recognised by CPID via dynamic loading (analogue to Marlin processors).

\section{CPID Module Library and Performance}

The current CPID framework contains a module library which covers basic training models and most PID observables that are used in detector models of future Higgs factories at the time of writing.
The training models implemented are a ROOT TMVA boosted decision tree (BDT) for signal and background and a multiclass BDT.
The input algorithms entail specific energy loss via \dEdx (with ILD full simulation) and dN/dx (cluster counting, based on Delphes parametrisation). For both, the distances of the measured value to the reference curves of the 5 detector-stable charged particles (5P: $e$, $\mu$, $\pi$, $K$, $p$) are used and can be scaled via a steering parameter to emulate a better or worse detector resolution than is intrinsic to the simulation.
Also available are time of flight (TOF) with an adjustable timing resolution, calorimeter cluster shapes (via Pandora ParticleFlow) and a dedicated LeptonID tool, which in itself combines cluster shapes with some \dEdx information for enhanced $e$ and $\mu$ identification.

\begin{figure}[!hbt]
  \centering
    \includegraphics[width=.4\textwidth,keepaspectratio=true]{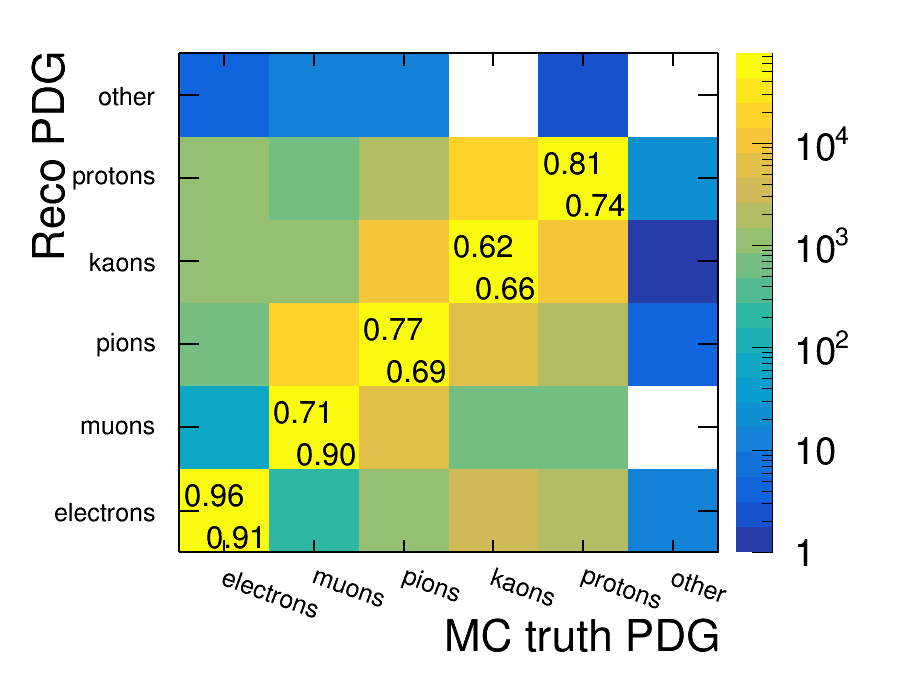}
		\hspace{1cm}
		\includegraphics[width=.4\textwidth,keepaspectratio=true]{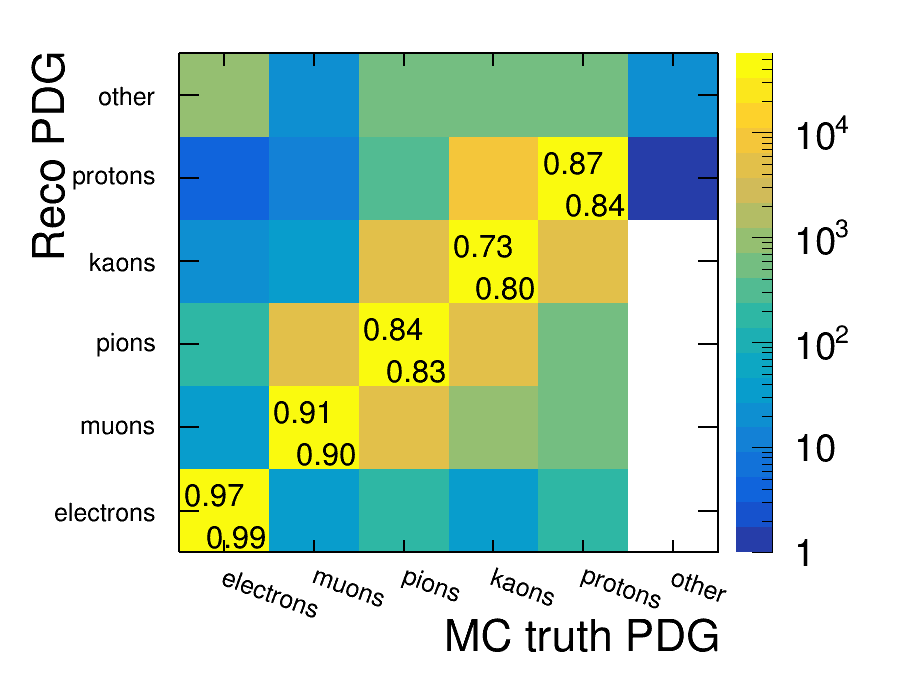}
    \caption{Comparison of PID at ILD with the current reconstruction (left) and with CPID (right), which has time of flight as additional input. The numbers indicate efficiency (upper) and purity (lower) of the corresponding diagonal bin.}
  \label{fig:ILD_PID}
\end{figure}

The performance of CPID is shown in \autoref{fig:ILD_PID}. The plots show confusion matrices of reconstructed PDG vs. true PDG for the 5P, based on isotropic single particle events in ILD. The numbers on the diagonal entries indicate the efficiency (upper) and purity (lower) for the corresponding species.
On the left, the ILD standard reconstruction is shown, which uses calorimeter cluster shapes as well as \dEdx information.
On the right, a reconstruction with CPID using the multiclass BDT model is shown. While CPID with the same input as the standard tool would result in a comparable plot with also comparable efficiency and purity numbers, here also the TOF observable (assuming 30 ps timing resolution per particle) was used, which leads to a significantly improved confusion matrix and higher efficiency and purity values for all species.
The point is, that while it would have been a significant effort to implement TOF into the standard tool, in CPID it can be included by adding one line to the steering file once a module has been created.

\begin{wrapfigure}{l}{.4\textwidth}
  \centering
    \includegraphics[width=.4\textwidth,keepaspectratio=true]{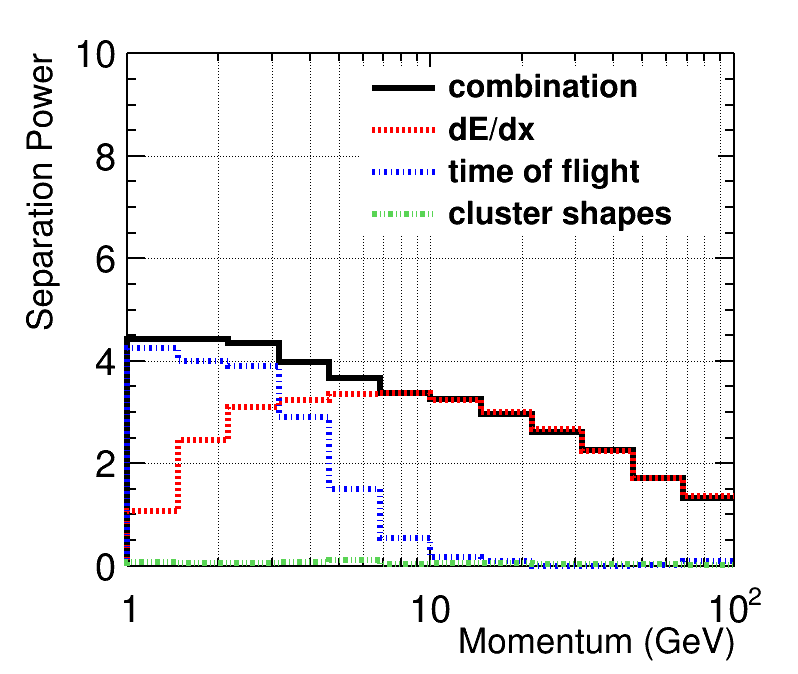}
    \caption{$\pi/K$ separation power at ILD evaluated with CPID and using different PID inputs.}
  \label{fig:ILD_pikaSP}
\end{wrapfigure}

In addition, \autoref{fig:ILD_pikaSP} shows the $\pi/K$ separation power, again based on isotropic single particles in ILD.
Here, by switching on different modules in the steering file, several training and inference runs of CPID using a signal/background BDT were performed with different individual input algorithms active and once with all three active in combination, and in each case with 12 momentum bins between 1 and 100 GeV.
The curve for \dEdx shows the typical shape: a minimum at the blind spot where the Bethe-Bloch curves of $\pi$s and $K$s overlap, a maximum at a few GeV and then a decline to large momenta.
The TOF curve has a maximum at low momenta and falls rapidly, reaching 0 at about 10 GeV.
The clusters shapes, used only for a cross-check, have no separation power between $\pi$s and $K$s, as expected.
The combination is close to a root-mean-squared of the individual contributions, again as expected for uncorrelated inputs.

\section{Conclusion and Outlook}

PID is increasingly used in analyses for future colliders, not least in the variety of flavour physics that emerged with the proposals of circular colliders.
In particular strange tagging has become an addition to the increasing numbers of flavour taggers, for which $K$ ID is crucial.
One analysis which pioneered this usage to determine limits on the strange Yukawa coupling is shown in \cite{Higgs_ssbar}.
CPID provides a coherent framework and evaluation for PID, which allows to compare different settings of detector models with varying resolutions or technologies in order to optimise them, but also to compare entirely different detectors in a consistent way inside the Key4hep framework.
In the future, more modules will be implemented, in particular an input algorithm for a Ring Imaging Cherenkov detector (RICH), as well as more ROOT-based TMVAs and an interface to 'external' machine learning approaches.
The implementation of CPID in ILD-related analyses and flavour tagging is ongoing.
The software is available at \cite{MarlinReco}.

\section{Acknowledgments}
In this study we used the National Analysis Facility (NAF) and would like to
thank Grid computational resources operated at Deutsches Elektronen-Synchrotron (DESY), Hamburg, Germany. We thankfully acknowledge the support by the Deutsche Forschungsgemeinschaft (DFG, German Research Foundation) under Germany’s Excellence Strategy EXC 2121 "Quantum Universe" 390833306.

\bibliographystyle{JHEP}
\bibliography{References_reduced}
\end{document}